\documentclass[12pt]{article}
\usepackage{amsmath,amsfonts,amssymb,cite}
\newcommand{\be}{\begin{equation}}
\newcommand{\ee}{\end{equation}}
\newcommand{\ba}{\begin{eqnarray}}
\newcommand{\ea}{\end{eqnarray}}

\newcommand{\la}[1]{\label{#1}}

\def\gl#1{(\ref{#1})}

\date{}
\begin{document}
\title{Polynomial supersymmetry for matrix\\ Hamiltonians: proofs}
\author{
A.V. Sokolov\footnote{E-mail: avs\_avs@rambler.ru, andrey.sokolov@spbu.ru}
\\{\ }\\
{\it Saint Petersburg State University, SPbSU,}\\ {\it 7/9 Universitetskaya nab., St. Petersburg 199034, Russia}
\\ {\it and Military Institute (Engineering-Technical),}\\ {\it 22 Zakharyevskaya street, St. Petersburg 191123, Russia}} 

\maketitle \abstract{In this paper we continue studying of matrix $n\times n$ linear differential intertwining operators. The problems of minimization and of reducibility of matrix intertwining operators are considered and criterions of weak minimizability and of regular reducibility are proven. It is shown that in contrast to the scalar case $n=1$ there are for any $n\geqslant2$ regularly absolutely irreducible matrix intertwining operators of any order $N\geqslant2$, {\it i.e.} operators which cannot be factorized into a product of a matrix intertwining operators of lower orders even with a pole singularity(-ies) into coefficients. The theorem is proven on existence for any matrix intertwining operator $Q_N^-$ of the order $N$ a matrix differential operator $Q_{N'}^+$ of other, in general, order $N'$ that intertwines the same Hamiltonians $H_+$ and $H_-$ as $Q_N^-$ in the opposite direction and such that the products $Q_{N'}^+Q_N^-$ and $Q_N^-Q_{N'}^+$ are identical polynomials of $H_+$ and $H_-$ respectively. Polynomial algebra of supersymmetry corresponding to this case is constructed.}

\section{Introduction}

The present paper is devoted to continuation of investigation of supersymmetry with matrix Hamiltonians from \cite{sokolov13,anso15,sokolov15,anso16}. The papers \cite{sokolov13,anso15} contain brief results without proofs on constructing of a matrix intertwining operator in terms of transformation vector-functions (including the case with formal associated vector-functions of initial Hamiltonian), on weak and strong minimizability and on reducibility of a matrix intertwining operator, on possibility to construct polynomial supersymmetry in some partial case and on constructing of matrix Hamiltonians with a given symmetry matrix. In \cite{sokolov15} general constructions of first-order and higher-order matrix $n\times n$ differential operators that intertwine matrix non-Hermitian, in general, Hamiltonians were described and founded. We consider in this paper the problems of weak minimization and of regular reducibility of matrix intertwining operators with proofs and show that polynomial algebra of supersymmetry can be constructed for arbitrary matrix intertwining operator (see review of papers on these problems in  \cite{sokolov13,sokolov15}). Some results of this paper were briefly presented earlier without proofs in \cite{sokolov12,sokolov13}. The paper \cite{anso16} contains some results on extended algebra of supersymmetry with matrix Hamiltonians.

This paper is organized as follows. In the Sec. 2 the basic notation is given. The Sec.~3 contains Theorem~1 on factorization of a matrix differential intertwining operator into a product of matrix differential intertwining operators of lower orders. This theorem takes an important part in investigation of polynomial supersymmetry with matrix intertwining operators and as well in studying of minimizability and of reducibility of such operators. The Sec. 4 comprises Theorem 2 on sufficient condition of existence for a matrix differential intertwining operator $Q_N^-$ of the order $N$, that intertwines a matrix Hamiltonians $H_+$ and $H_-$, a matrix differential operator $Q_N^+$ of the same order $N$ that intertwines the same Hamiltonians in the opposite direction and such that the products $Q_N^+Q_N^-$ and $Q_N^-Q_N^+$ are identical polynomials of $H_+$ and $H_-$ respectively. Polynomial algebra of supersymmetry is constructed of these $H_+$, $H_-$, $Q_N^+$ and $Q_N^-$ in the Sec. 4 as well. The Sec. 5 includes definitions of minimizability in different senses and, in particular, of weak minimizability of a matrix differential intertwining operator and criterion of weak minimizability for such operator (Theorem 3). Definitions of regularly and singularly reducible and irreducible and of regularly absolutely irreducible matrix differential intertwining operators, criterion of regular reducibility for such operators (Theorem 4) and proof of existence of regularly absolutely irreducible matrix $n\times n$ differential intertwining operators of the $N$-th order for any $n\geqslant2$ and $N\geqslant2$ are presented in the Sec. 6. The Sec. 7 contains Theorem 5 on existence for any matrix intertwining operator $Q_N^-$ of the order $N$ a matrix operator $Q_{N'}^+$ of the order $N'$ different, in general, from $N$ that intertwines the same Hamiltonians $H_+$ and $H_-$ as $Q_N^-$ in the opposite direction and such that the product $Q_{N'}^+Q_N^-$ is a polynomial of $H_+$. Moreover, it is shown in the Sec. 7 that the product $Q_N^-Q_{N'}^+$ is the same polynomial of $H_-$ and polynomial algebra of supersymmetry is constructed from $H_+$, $H_-$, $Q_{N'}^+$ and $Q_N^-$. Conclusion contains list of problems which can be considered in the following papers.

\section{Basic definitions and notation \label{sect2}}

We consider two matrix Hamiltonians of Schr\"odinger form,
\begin{equation}H_+=-I_n\partial^2+V_+(x),\quad H_-=-I_n\partial^2+V_-(x),\qquad\partial\equiv{d\over{dx}},\la{h+h-2.1}\end{equation} 
defined on the entire axis, where $I_n$ is the identity matrix $n\times n$, $n\in\Bbb N$, and all entries of the matrix $n\times n$ potentials $V_+(x)$ and $V_-(x)$ are supposed to be sufficiently smooth and, in general, complex-valued functions. The Hamiltonians $H_+$ and $H_-$ are
intertwined by a matrix $n\times n$ linear differential operator $Q_N^-$ of the order $N$, $N\in{\Bbb N}\cup\{0\}$, so that
\begin{equation} Q_N^-H_+=H_-Q^-_N,\qquad
Q_N^-=\sum\limits_{j=0}^NX^-_j(x)\partial^j,\la{splet}\end{equation}
and all entries of the matrix coefficients $X^-_j(x)$, $j=0$, \dots, $N$ are also supposed to be sufficiently smooth and, in
general, complex-valued functions. It follows \cite{sokolov13} from (\ref{splet}) that the coefficient $X^-_N(x)$ is a constant
matrix and
\begin{equation}V_-(x)=X^-_NV_+(x)(X^-_N)^{-1}+2X^{-\,\prime}_{N-1}(x)(X^-_N)^{-1}\qquad\text{if}\qquad\det X^-_N\ne0.\la{vmp2.1}\end{equation}
We assume in what follows that $\det X^-_N\ne0$. 

Since the kernel of $Q_N^-$ on the strength of \gl{splet} is an invariant subspace with respect to~$H_+$: 
\[H_+\ker Q_N^-\subset\ker Q_N^-.\] 
so for any basis $\Phi^-_1(x)$, \dots, $\Phi^-_d(x)$, $d=\dim\ker Q_N^-=nN$ in this subspace there is a constant
square $d\times d$ matrix ${\bf T}^+\equiv\|T^+_{ij}\|$ for which the following relations are valid,
\begin{equation}H_+\Phi^-_i=\sum_{j=1}^dT^+_{ij}\Phi^-_j,\qquad i=1,\ldots,d.
\la{tm}\end{equation} 
It should be noted that the Wronskian of all elements of any basis in the kernel of $Q_N^-$ is nonvanishing on the entire axis in view of Liouville -- Ostrogradsky formula.

We define the {\it matrix} $\bf T$ of an intertwining operator as a matrix that is built for this operator similarly as the matrix ${\bf T}^+$ was constructed for $Q_N^-$. Herein the basis in the kernel of the intertwining operator is not specified if we consider the spectral characteristics only of the matrix $\bf T$ or, that the same, the spectral characteristics of the restriction of the corresponding Hamiltonian to the kernel of the intertwining operator in question (cf. with~\gl{tm}).

A basis in the kernel of an intertwining operator that provides a normal (Jordan) form to the matrix $\bf T$ of this operator is defined as a {\it canonical basis}. Elements of a canonical basis are defined as {\it transformation vector-functions}.

In the cases where a normal form of the matrix $\bf T$ of an intertwining operator comprises block(s) of order(s) greater than one the corresponding canonical basis includes besides formal vector-eigenfunction(s) of the corresponding Hamiltonian its formal associated vector-function(s) as well. The latter are defined \cite{naim} as follows.

A vector-function $\Phi_{m,i}(x)$ is defined as a {\it formal associated vector-function of the or\-der $i$} of a matrix $n\times n$ Hamiltonian
$H=-I_n\partial^2+V(x)$ for a spectral value $\lambda_m$ if
\[(H-\lambda_mI_n)^{i+1}\Phi_{m,i}\equiv 0,\qquad(H-\lambda_mI_n)^{i}\Phi_{m,i}\not\equiv0,\] 
and the word ``formal'' underlines herein that the vector-function $\Phi_{m,i}(x)$ is
not necessarily belongs to $L^2({\Bbb R},{\Bbb C}^n)$ (not necessarily normalizable). A formal associated
vector-function of zero order $\Phi_{m,0}(x)$ in this case is a formal vector-eigenfunction of $H$.

A finite or infinite set of vector-functions $\Phi_{m,i}(x)$, $i=0$, 1, 2, \dots\, is called a {\it chain} of formal associated vector-functions of a matrix $n\times n$ Hamiltonian $H=-I_n\partial^2+V(x)$ for a spectral value $\lambda_m$ if \[H\Phi_{m,0}=\lambda_m\Phi_{m,0},\quad\Phi_{m,0}(x)\not\equiv0,
\qquad (H-\lambda_mI_n)\Phi_{m,i}=\Phi_{m,i-1},\quad i=1,2,3,\ldots\,.\] It is evident that $\Phi_{m,i}(x)$ in this case is a formal associated vector-function of $i$-th order of the Hamiltonian $H$ for the spectral value $\lambda_m$, $i=0$, 1, 2, \dots\,.

A chain $\Psi_{m,l}^-(x)$, $l=0$, 1, 2, \dots\, of formal associated vector-functions of the Hamiltonian $H_+$ for a spectral value $\lambda_m$ in view of the equalities 
\[(H_--\lambda_mI_n)Q_N^-\Psi_{m,l}^-=Q_N^-(H_+-\lambda_mI_n)\Psi_{m,l}^-=
Q_N^-\Psi_{m,l-1}^-,\] 
\begin{equation}l=0,1,2,\ldots,\qquad \Psi_{m,-1}^-(x)\equiv0,\la{map}\end{equation} 
that take place due to \gl{splet}, is mapped by $Q_N^-$ into a chain of formal associated vector-functions of the Hamiltonian $H_-$ for the same spectral value $\lambda_m$ with possible exception of some number of vector-functions $Q_N^-\Psi_{m,l}^-$ with lower numbers which can be identical zeroes. It is clear in view of \gl{map} that if $Q_N^-\Psi_{m,l_0}^-\equiv0$ for some $l_0$ then $Q_N^-\Psi_{m,l}^-\equiv0$ for any $l<l_0$ and if $Q_N^-\Psi_{m,l_0}^-\not\equiv0$ for some $l_0$ then $Q_N^-\Psi_{m,l}^-\not\equiv0$ for any $l>l_0$. Thus, if $l_0$ is a minimal number such that $Q_N^-\Psi_{m,l_0}^-\not\equiv0$ then one can represent the arising chain of formal associated vector-functions  of $H_-$ in the form \[\Psi_{m,l}^+(x)=Q_N^-\Psi_{m,l+l_0}^-(x), \qquad l=0,1,2,\ldots\,.\] 

\section{Factorization of a matrix intertwining operator\la{factsin}}

The following theorem on factorization of a matrix differential intertwining operator into a product of a matrix intertwining operators of lower orders takes an important part in investigation of polynomial supersymmetry with matrix intertwining operators and in studying of minimizability and reducibility of such operators (see below sections~\ref{secpol}~--~\ref{secpolgen}).\\

\noindent {\bf Theorem 1} (on a singular factorization of a matrix intertwining operator).\\ {\it Suppose that
\renewcommand{\labelenumi}{\rm{(\theenumi)}}
\begin{enumerate}
\item a matrix $n\times n$ Hamiltonians $H_+$ and $H_-$ of the form {\rm\gl{h+h-2.1}} are intertwined by a matrix $n\times n$ linear differential operator of $N$-th order $Q_N^-$ of the form {\rm\gl{splet}} with constant nondegenerate matrix coefficient $X_N^-$ at $\partial^N$ in accordance with {\rm\gl{splet};}
\item $\Phi_l^-(x)\equiv(\varphi_{l1}^-(x),\ldots,\varphi_{ln}^-(x))^t$, $l=1$, \dots, $nN$ are elements of a canonical basis in $\ker Q_N^-$ renumbered so that the set
\[\Phi_l^-(x),\qquad l=1,\ldots,nj\] for any $j=1$, \dots, $N-1$ can be divided into a chains of formal associated vector-functions of 
$H_+;$
\item the Wronskian $W_j(x)$ is defined by the equality
\[W_j(x)=\begin{vmatrix}\varphi^-_{11}&\dots&\varphi^-_{1n}& \varphi^{-\prime}_{11}&\dots&\varphi^{-\prime}_{1n}&\ldots&(\varphi^-_{11})^{(j-1)}&\dots&(\varphi^-_{1n})^{(j-1)}\\
\varphi^-_{21}&\dots&\varphi^-_{2n}&\varphi^{-\prime}_{21}&\dots&\varphi^{-\prime}_{2n}&\ldots&
(\varphi^-_{21})^{(j-1)}&\dots&(\varphi^-_{2n})^{(j-1)}\\
\vdots&\ddots&\vdots&\vdots&\ddots&\vdots&\ddots&\vdots&\ddots&\vdots\\
\varphi^-_{nj,1}\!\!&\dots&\!\!\varphi^-_{nj,n}&\varphi^{-\prime}_{nj,1}\!\!&\dots&\!\!\varphi^{-\prime}_{nj,n}\!\!&\ldots&\!\!(\varphi^-_{nj,1})^{(j-1)}\!\!&\dots&\!\!(\varphi^-_{nj,n})^{(j-1)}\end{vmatrix},\]

\[j=1,\ldots,N;\]

\item $j_m$, $m=1$, \ldots, $M$ is monotonically increasing sequence of natural numbers such that $j_M=N$ and \[W_{j_m}(x)\not\equiv0,\qquad m=1,\ldots,M-1;\]
\item the number $N_m$ is defined by the equality \[N_m=j_m-j_{m-1},\qquad m=1,\ldots,M, \quad j_0\equiv0.\]
\end{enumerate}
Then there exist the matrix $n\times n$ linear differential operators $Q_{N_m,m}^-$, $m=1$, \dots, $M$ such that$:$
\renewcommand{\labelenumi}{\rm{(\theenumi)}}
\begin{enumerate}
\item the operator $Q_{N_m,m}^-$ is an operator of $N_m$-th order of the form
\[Q_{N_m,m}^-=\sum_{j=0}^{N_m}X_{j,m}^-(x)\partial^j,\qquad X_{N_m,m}^-(x)\equiv I_n,\quad m=1,\ldots,M,\]
where $X_{j,m}^-(x)$ is an $n\times n$ matrix-valued function, any entry of which is, in general, complex-valued even in the case where all entries of $X_0^-(x)$, \ldots, $X_{N-1}^-(x)$, $X_N^-$, $V_+(x)$ and $V_-(x)$ are real-valued, and possesses, in general, by a pole$($s$)$, $j=0$, \dots, $N_m-1$, $m=1$, \dots, $M;$
\item the following relations hold,
\[Q_{N_m,m}^-\cdot\ldots\cdot Q_{N_1,1}^-\Phi_l^-=0,\qquad Q_{N_{m-1},m-1}^-\cdot\ldots\cdot Q_{N_1,1}^-\Phi_l^-\not\equiv0,\]
\[l=nj_{m-1}+1,\ldots,nj_m, \quad m=1,\ldots,M;\]
\item the Wronskian of the vector-functions $Q_{N_{m-1},m-1}^-\cdot\ldots\cdot Q_{N_1,1}^-\Phi_l^-$, $l\!=\!nj_{m-1}\!+\!1$, \dots, $nj_m$ is different from identical zero and the determinant representation $($see {\rm(41), (42)} in {\rm\cite{sokolov15}}$)$
for the operator $Q_{N_m,m}^-$ in terms of these vector-functions takes place, $m\!=\!1$, \dots,~$M;$
\item the following factorization of $Q_N^-$ is valid,
\begin{equation}Q_N^-=X_N^-Q_{N_M,M}^-\cdot\ldots\cdot Q_{N_1,1}^-;\la{factor6}\end{equation}
\item there are the matrix $n\times n$ Hamiltonians $H_m$, $m=1$, \dots, $M$ of Schr\"odinger form, \begin{equation}H_m=-I_n\partial^2+V_m(x),\qquad m=1,\ldots,M\la{hm6}\end{equation} such that$:$
\renewcommand{\labelenumii}{\rm{(\theenumii)}}
\begin{enumerate}
\item the potential $V_m(x)$ is an $n\times n$ matrix-valued function, any entry of which is, in general, complex-valued even in the case where all entries of $X_0^-(x)$,~\ldots, $X_{N-1}^-(x)$, $X_N^-$, $V_+(x)$ and $V_-(x)$ are real-valued, and possesses, in general, by a pole$($s$)$, $m=1$, \dots, $M-1$ and the potential $V_M(x)$ is an $n\times n$ matrix-valued function, any entry of which is smooth $($without pole$($s$))$ and, in general, complex-valued but real-valued in the case where all entries of $V_-(x)$ and $X_N^-$ are real-valued$;$
\item the following chain relations hold,
\[Q_{N_m,m}^-H_{m-1}=H_mQ_{N_m,m}^-,\qquad m=1,\ldots, M,\qquad\qquad\qquad\qquad\qquad\,\,\,\]
\begin{equation}\qquad\qquad\qquad\qquad\qquad\qquad\qquad\qquad H_0\equiv H_+,\qquad X_N^-H_M=H_-X_N^-;\la{cepsplet6}\end{equation}
\item the potential $V_m(x)$ can be found either in the chain with the help of the equalities
\[V_m(x)=V_{m-1}(x)+2X_{N_m-1,m}^{-\prime}(x),\qquad m=1,\ldots,M,\quad V_0(x)\equiv V_+(x)\]
or the equalities
\[V_m(x)=V_{m+1}(x)-2X^{-\prime}_{N_{m+1}-1,m+1}(x),\!\qquad\qquad\qquad\qquad\qquad\qquad\qquad\] 
\begin{equation}\qquad\qquad\qquad\qquad\qquad\! m=M-1,\ldots,1,\quad V_M(x)=(X_N^-)^{-1}V_-(x)X_N^-\la{vmnaz6}\end{equation}
or in terms of $V_+(x)$ and the coefficient of $Q^-_{N_m,m}\cdot\ldots\cdot Q_{N_1,1}^-$ at $\partial^{j_m-1}$ represented with the use of $\Phi_l^-(x)$, $l=1$, \dots, $nj_m$ and {\rm(42)} from {\rm\cite{sokolov15}} with the help of the equality {\rm\gl{vmp2.1}} rewritten for the considered case, $m=1$, \dots, $M$ or in terms of $V_-(x)$ and the coefficient of $X_N^-Q^-_{N_M,M}\cdot\ldots\cdot Q_{N_{m+1},m+1}^-$ at $\partial^{N-j_m-1}$ represented with the use of $Q^-_{N_m,m}\cdot\ldots\cdot Q_{N_1,1}^-\Phi_l^-(x)$, $l=nj_m+1$, \dots, $nN$ and {\rm(42)} from {\rm\cite{sokolov15}} with the help of the equality {\rm\gl{vmp2.1}} rewritten for the considered case, $m\!=\!1$, \!\dots,\! $M;$
\end{enumerate}
\item the vector-functions $Q_{N_{m-1},m-1}^-\cdot\ldots\cdot Q_{N_1,1}^-\Phi_l^-$, $l=nj_{m-1}+1$, \dots, $nj_m$ after appropriate renumbering form a canonical basis in $\ker Q_{N_m,m}^-$, $m=1$, \dots, $M;$ moreover, the eigenvalue of the matrix $\bf T$ of the operator $Q_{N_m,m}^-$ corresponding to $Q_{N_{m-1},m-1}^-\cdot\ldots\cdot Q_{N_1,1}^-\Phi_l^-$ is equal to the eigenvalue of the matrix $\bf T$ of the operator $Q_{N}^-$ corresponding to $\Phi_l^-$, $l=nj_{m-1}+1$, \dots, $nj_m$, $m=1$, \dots, $M$.
\end{enumerate}}

\vskip1pc

{\bf Proof.} Let us consider the Hamiltonian $H_M=(X_N^-)^{-1}H_-X_N^-$. It is evident that this Hamiltonian is of Schr\"odinger form \gl{hm6} with the potential $V_M(x)$ from \gl{vmnaz6} and is intertwined with the Hamiltonian $H_-$ in accordance with \gl{cepsplet6} and with the Hamiltonian $H_+$ as follows, \begin{equation}\tilde Q_N^-H_+=H_M\tilde Q_N^-,\qquad\tilde Q_N^-=(X_N^-)^{-1}Q_N^-,\la{int16}\end{equation} where the $n\times n$ matrix coefficient of $\tilde Q_N^-$ at $\partial^N$ is equal to $I_n$. Moreover, the potential $V_M(x)$ satisfies the paragraph (a) of the statement (5) and $\ker\tilde Q_N^-=\ker Q_N^-$, wherefrom it follows that the set $\Phi_l^-(x)$, $l=1$, \dots, $nN$ is a canonical basis in $\ker \tilde Q_N^-$ as well.

In view of the condition $W_{j_{M-1}}(x)\not\equiv0$ we can construct (see section 4 of \cite{sokolov15}) the $n\times n$ matrix linear differential operator of $j_{M-1}$-th order $\hat Q_{j_{M-1},M-1}^-$ with the coefficient $I_n$ at $\partial^{j_{M-1}}$ in terms of vector-functions $\Phi_l^-(x)$, $l=1$, \dots, $nj_{M-1}$ with the help of the formulae analogous to (41) and (42) from \cite{sokolov15}, so that there is the Hamiltonian $H_{M-1}$ of Schr\"odinger form \gl{hm6} that satisfies the paragraph (a) of the statement (5) and the following intertwining takes place, 
\begin{equation}\hat Q_{j_{M-1},M-1}^-H_+=H_{M-1}\hat Q_{j_{M-1},M-1}^-.\la{int26}\end{equation} Let us show that there is an $n\times n$ matrix linear differential intertwining operator $Q_{N_M,M}^-$ with the coefficient $I_n$ at $\partial^{N_M}$ such that \begin{equation}\tilde Q_N^-= Q_{N_M,M}^-\hat Q_{j_{M-1},M-1}^-\la{factq6}\end{equation} and the intertwining \gl{cepsplet6} for $m=M$ is valid.

To prove the possibility of separation of $\hat Q_{j_{M-1},M-1}^-$ from the right-hand side of $\tilde Q_N^-$ we consider a basis 
$\Psi_l(x)\equiv \big(\psi_{l1}(x),\ldots,\psi_{ln}(x)\big)^t$, $l=1$, \dots, $nj_{M-1}$ in $\ker\hat Q_{j_{M-1},M-1}^-$ and note that $\ker\hat Q_{j_{M-1},M-1}^-\subset\ker\tilde Q_N^-$ since all elements of the basis $\Phi_l^-(x)$, $l=1$, \dots, $nj_{M-1}$ in $\ker\hat Q_{j_{M-1},M-1}^-$ belong to $\ker\tilde Q_N^-$ as well. Taking into account that the Wronskian of elements of any basis in the kernel of an $n\times n$ matrix linear differential operator with the coefficient $I_n$ at $\partial$ in the highest degree is different from identical zero and that the calculation of the Wronskian
\[\begin{vmatrix}
\psi_{11}&\dots&\psi_{1n}&\ldots&(\psi_{11})^{(j_{M-1}-1)}&\dots&(\psi_{1n})^{(j_{M-1}-1)}\\
\psi_{21}&\dots&\psi_{2n}&\ldots&(\psi_{21})^{(j_{M-1}-1)}&\dots&(\psi_{2n})^{(j_{M-1}-1)}\\
\vdots&\ddots&\vdots&\ddots&\vdots&\ddots&\vdots\\
\psi_{nj_{M-1},1}&\dots&\psi_{nj_{M-1},n}&\ldots&(\psi_{nj_{M-1},1})^{(j_{M-1}-1)}&\dots&(\psi_{nj_{M-1},n})^{(j_{M-1}-1)}
\end{vmatrix}\] by successive computations with the help of the last columns can be reduced to calculations of the Wronskians of the subsets consisting of $n$ elements of the considered basis in $\ker\hat Q_{j_{M-1},M-1}^-$, we come to the conclusion that there is a subset consisting of $n$ elements of the mentioned basis such that the Wronskian of elements of this subset is different from identical zero.
Consider the $n\times n$ matrix-valued function $Y(x)$, columns of which coincide with elements of this subset. One can separate the operator $I_n\partial-Y'(x)Y^{-1}(x)$ from the right-hand sides of $\tilde Q_N^-$ and $\hat Q_{j_{M-1},M-1}^-$ with the help of the following explicit formula \[\tilde Q_N^-\equiv\sum_{j=0}^N\tilde X_j^-(x)\partial^j=\bigg\{\sum_{j=1}^N\tilde X_j^-(x)\Big[\partial^{j-1}+\sum_{l=0}^{j-2}\partial^{j-l-2}Z_l(x)\Big]\bigg\}\big[I_n\partial-Y'(x)Y^{-1}(x)\big],\]
\[Z_l(x)=Z'_{l-1}(x)+Z_{l-1}(x)Y'(x)Y^{-1}(x),\qquad l=1,\ldots, N-2,\quad Z_0(x)=Y'(x)Y^{-1}(x)\] and the analogous formula for $\hat Q_{j_{M-1},M-1}^-$, which can be easily derived in the same way as (32) and (33) in \cite{anso03}. 
It is not hard to check that the nonzero elements from the set $[I_n\partial-Y'(x)Y^{-1}(x)]\Psi_l(x)$, $l=1$, \dots, $nj_{M-1}$ (there are only $n(j_{M-1}-1)$ such elements obviously) form a basis in the rest of $\hat Q_{j_{M-1},M-1}^-$ and belong to the kernel of the rest of $\tilde Q_N^-$. Hence, the kernel of the rest of $\hat Q_{j_{M-1},M-1}^-$ is a subset of the kernel of the rest of~$\tilde Q_N^-$. 
Proceeding further in the same way by induction, we obtain that the operator $\hat Q_{j_{M-1},M-1}^-$ can be factorized into a product of a matrix $n\times n$ linear differential operators of first order with the coefficients $I_n$ at $\partial$ and there is a matrix $n\times n$ linear differential operator of $N_m$-th order $Q_{N_M,M}^-$ with the coefficient $I_n$ at $\partial^{N_M}$ such that the factorization \gl{factq6} takes place. 

The fact that the operator $Q_{N_M,M}^-$ intertwines the Hamiltonians $H_{M-1}$ and $H_M$ in accordance with \gl{cepsplet6} is valid due to the chain \begin{eqnarray}H_MQ_{N_M,M}^-\hat Q_{j_{M-1},M-1}^-&=&H_M\tilde Q_N^-=\tilde Q_N^-H_+=Q_{N_M,M}^-\hat Q_{j_{M-1},M-1}^-H_+\nonumber\\&=&
Q_{N_M,M}^-H_{M-1}\hat Q_{j_{M-1},M-1}^-\nonumber\end{eqnarray} which follows from \gl{int16}, \gl{int26} and \gl{factq6}. The vector-functions $\hat Q_{j_{M-1},M-1}^-\Phi_l^-$, $l=nj_{M-1}+1$, \dots, $nN$ form obviously a basis in $\ker Q_{N_M,M}^-$. Hence, the Wronskian of these vector-functions is different from identical zero and the operator $Q_{N_M,M}^-$ can be represented in terms of these vector-functions with the help of the formulae analogous to (41) and (42) from \cite{sokolov15}. Moreover, it is evident in view of the last paragraph of section \ref{sect2} that the vector-functions $\hat Q_{j_{M-1},M-1}^-\Phi_l^-$, $l=nj_{M-1}+1$, \dots, $nN$ after appropriate renumbering form a canonical basis in $\ker Q_{N_M,M}^-$ and that the eigenvalue of the matrix $\bf T$ of the operator $Q_{N_M,M}^-$ corresponding to $\hat Q_{j_{M-1},M-1}^-\Phi_l^-$ is equal to the eigenvalue of the matrix $\bf T$ of the operator $Q_{N}^-$ corresponding to $\Phi_l^-$, $l=nj_{M-1}+1$, \dots, $nN$.

Proceeding further in the same way by induction, we conclude that the statements (1)~-- (4) and (6) and the paragraphs (a) and (b) of the statement (5) take place. The paragraph (c) of the statement (5) follows from the intertwinings \gl{cepsplet6} and from the formulae analogous to \gl{vmp2.1} in the corresponding cases. Theorem 1 is proved.\\

{\bf Remark 1.} In the conditions of Theorem 1 in the case where 
\[j_m=m,\quad m=1,\ldots, M=N\qquad\Rightarrow\qquad N_m=1,\quad m=1,\ldots,M=N\]
all intertwining operators $Q_{1,m}^-$, $m=1$, \dots, $N$ are intertwining operators of the first order and it is possible to present
in view of the results of section 3 from \cite{sokolov15}  the following additional chain relations as well as to simplify some formulae of Theorem 1:
\[Q^-_{1,j}H_{j-1}=H_jQ^-_{1,j},\qquad j=1,\dots,N,\qquad H_0\equiv H_+,\]
\[ H_j=Q_{1,j+1}^+Q^-_{1,j+1}+U_{0,j+1}(x)=Q^-_{1,j}Q^+_{1,j}+U_{0,j}(x),\qquad j=1,\dots,N-1,\]
\[H_0=Q^+_{1,1}Q^-_{1,1}+U_{0,1}(x),\qquad H_N=Q^-_{1N}Q^+_{1N}+U_{0,N}(x),\]
\[[U_{0,j}(x),Q_{1,j}^-]=0,\qquad j=1,\ldots,N,\]
\[Q_{1,j}^-\!\equiv\! I_n\partial\!+\!X^-_{0,j}(x),\quad Q_{1,j}^+\!\mathop{=}\limits^{\text{def}}\!-I_n\partial\!+\!X^-_{0,j}(x),
\quad H_j\equiv-I_n\partial^2+V_j(x), \qquad j\!=\!1,\ldots,N,\]

\[U_{0,j}(x)=V_{j-1}(x)-(X^-_{0,j}(x))^2+X^{-\prime}_{0,j}(x),\qquad V_j(x)=(X^-_{0,j}(x))^2+X^{-\prime}_{0j}(x)+U_{0,j}(x)\]
\[\equiv V_{j-1}(x)+2X^{-\prime}_{0,j}(x),\qquad j=1,\ldots,N,\qquad V_0(x)\equiv V_+(x),\]

\[U_{0,j}(x)=V_{j}(x)-(X^-_{0,j}(x))^2-X^{-\prime}_{0,j}(x),\qquad V_{j-1}(x)=(X^-_{0,j}(x))^2-X^{-\prime}_{0j}(x)+U_{0,j}(x)\]
\[\equiv V_{j}(x)-2X^{-\prime}_{0,j}(x),\qquad j=N,\ldots,1,\qquad V_N(x)\equiv(X_N^-)^{-1} V_-(x)X_N^-,\]
where any entry of any of $n\times n$ matrix-valued functions $U_{0,j}(x)$, $j=1$, \dots, $N$ is complex-valued, in general, even in the case where all entries of $X_0^-(x)$, \ldots, $X_{N-1}^-(x)$, $X_N^-$, $V_+(x)$ and $V_-(x)$ are real-valued, and possesses, in general, by a pole$($s$)$.\\

{\bf Remark 2.} One can reformulate Theorem 1 so that it contains the following factorization of the intertwining operator $Q_N^-$ instead of \gl{factor6}, \[Q_N^-=\tilde Q_{N_M,M}^-\cdot\ldots\cdot\tilde Q_{N_1,1}^-X_N^-,\] where $\tilde Q_{N_m,m}^-$ is the matrix $n\times n$ linear differential intertwining operator of the $N_m$-th order with the coefficient $I_n$ at $\partial^{N_m}$, $m=1$, \dots, $M$. It is evident that one can find these operators as follows, \[\tilde Q_{N_m,m}^-=X_N^-Q_{N_m,m}^-(X_N^-)^{-1},\qquad m=1,\ldots,M\]
and the corresponding intermediate Hamiltonians $\tilde H_m$, $m=0$, \dots, $M-1$ of Schr\"odinger form that satisfy the chain relations 
\[\tilde Q_{N_m,m}^-\tilde H_{m-1}=\tilde H_m\tilde Q_{N_m,m}^-,\qquad m=1,\ldots,M,\qquad\qquad\qquad\qquad\qquad\,\,\,\]
\[\qquad\qquad\qquad\qquad\qquad\qquad\qquad\qquad\,\,\,\tilde H_M\equiv H_-,\qquad X_N^-H_+=\tilde H_0X_N^-\] can be found with the help of the equalities 
\[\tilde H_m=X_N^-H_m(X_N^-)^{-1},\qquad m=0,\ldots,M.\]

%
%
%

\section{A partial case with polynomial SUSY \la{secpol}}

The following theorem contains the sufficient condition (different from the trivial conditions of the Sec. 2.2 of \cite{sokolov15}) which provides for a matrix $n\times n$ linear differential operator $Q_N^-$ intertwining a matrix Hamiltonians $H_+$ and $H_-$ in accordance to \gl{splet} existence of the conjugate in some sense matrix $n\times n$ linear differential operator $Q_N^+$ of $N$-th order that intertwines the same Hamiltonians $H_+$ and $H_-$ as $Q_N^-$ in the opposite direction. Moreover, by this theorem the products $Q_N^+Q_N^-$ and $Q_N^-Q_N^+$ are identical polynomials of the Hamiltonians $H_+$ and $H_-$ respectively. The latter allows us to construct  polynomial algebra of supersymmetry from $H_+$, $H_-$, $Q_N^+$ and~$Q_N^-$.\\

\noindent{\bf Theorem 2} (sufficient condition of existence of ``conjugate'' intertwining operator $Q_N^+$ for a given intertwining operator $Q_N^-$ and of polynomial supersymmetry).
{\it Suppose that
\renewcommand{\labelenumi}{\rm{(\theenumi)}}
\begin{enumerate}
\item the conditions of Theorem $1$ and Remark $1$ takes place$;$
\item $T^+$ is the matrix $T$ of the operator $Q_N^-;$
\item for any eigenvalue of the matrix $T^+$ there are $n$ $($and no more$)$ Jordan blocks corresponding to this eigenvalue in a Jordan form of $T^+$ and, moreover, all these blocks $($for any fixed eigenvalue$)$ have identical sizes$;$
\item $j_m=m$, $m=1$, \dots, $M=N;$
\item all vector-functions $\Phi_l^-(x)$, $l=n(m-1)+1$, \dots, $nm$ correspond to the same eigenvalue $\lambda_m$ of the matrix $T^+$ and are associated vector-functions of $H_+$ of the same order, $m=1$, \dots, $N;$
\item the polynomial ${\cal P}_N(\lambda)$ is defined by the equality ${\cal P}_N(\lambda)=\prod\limits_{m=1}^N(\lambda-\lambda_m);$
\item $X_{0,m}^-(x)$ is the matrix $n\times n$ coefficient of $Q_{1,m}^-$ at $\partial^0$, so that $Q_{1,m}^-=I_n\partial+X_{0,m}^-(x)$, $m=1$, \dots, $N;$
\item the operator $Q_{1,m}^+$ is defined by the equality $Q_{1,m}^+=-I_n\partial+X_{0,m}^-(x)$, $m=1$, \dots, $N;$
\item the operator $Q_N^+$ is defined by the the equality $Q_N^+=Q_{1,1}^+\cdot\ldots\cdot Q_{1,N}^+(X_N^-)^{-1}.$
\end{enumerate}
Then
\renewcommand{\labelenumi}{\rm{(\theenumi)}}
\begin{enumerate}
\item all vector-functions $Q_{1,m-1}^-\cdot\ldots\cdot Q_{1,1}^-\Phi_l^-$, $l=n(m-1)+1$, \dots, $nm$ are formal vector-eigenfunctions of the Hamiltonian $H_{m-1}$, $m=1$, \dots, $N;$
\item the $n\times n$ matrix-valued functions $U_{0,m}(x)$, $m=1$, \dots, $N$ take the following form \[U_{0,m}(x)=\lambda_mI_n,\qquad m=1,\ldots,N;\]
\item the following chain relations hold, \begin{equation}H_{m-1}Q_{1,m}^+=Q_{1,m}^+H_m,\qquad m=1,\ldots,N,\qquad H_0\equiv H_+,\la{int71}\end{equation} \begin{equation}H_N(X_N^-)^{-1}=(X_N^-)^{-1}H_-;\la{int72}\end{equation}
\item the operator $Q_N^+$ can be represented in the form \[Q_N^+=\sum_{j=1}^NX_j^+(x)\partial^j,\] where $X_N^+=(-1)^N(X_N^-)^{-1}$ and $X_j^+(x)$ is an $n\times n$ matrix-valued function, all elements of which are smooth $($without pole$($s$))$, $j=1$, \dots, $N-1;$
\item if all elements of all matrix-valued functions $V_+(x)$, $X_0^-(x)$, \dots, $X_{N-1}^-(x)$ and of the matrix $X_N^-$ are real-valued then all elements of all matrix-valued functions $X_0^+(x)$, \dots, $X_{N-1}^+(x)$ and of the matrix $X_N^+$ are real-valued as well$;$ otherwise all elements of the matrix-valued functions $X_0^+(x)$, \dots, $X_{N-1}^+(x)$ and of the matrix $X_N^+$ are complex-valued, in general$;$
\item the operator $Q_N^+$ intertwines the Hamiltonians $H_+$ and $H_-$, so that \[H_+Q_N^+=Q_N^+H_-;\]
\item the operator $Q_N^+$ does not depend on the order of numbering of the numbers $\lambda_m$, $m=1$, \dots, $N$ and of the order of numbering of  the vector-functions $\Phi_l^-(x)$, $l=1$, \dots, $nN$ corresponding to a fixed numeration of $\lambda_m$, $m=1$, \dots, $N;$
\item the following equalities hold, \begin{equation}Q_N^+Q_N^-={\cal P}_N(H_+),\qquad Q_N^-Q_N^+={\cal P}_N(H_-);\la{prod7}\end{equation}
\item the normal $($Jordan$)$ forms of the matrices $T$ of the operators $Q_N^+$ and $Q_N^-$ are identical up to permutation of Jordan blocks.
\end{enumerate}}

\vskip1pc

{\bf Proof.} The statements (1) and (2) follow in view of the Sec. 2.3 of \cite{sokolov15} from Theorem~1, Remark~1 and the representations analogous to (43) from \cite{sokolov15} for the matrix-valued functions $U_{0,m}(x)$, $m=1$, \dots, $N$. The intertwining relations \gl{int71} are analogous to (20) from \cite{sokolov15} and take place in view of Remark~1 by virtue of the validity of the conditions analogous to (21) from \cite{sokolov15} which follow from the statement~(2). The intertwining relation \gl{int72} holds due to \gl{cepsplet6} with $M=N$. 
The statement (6) is a corollary of the statement (3). The statement (8) follows from the factorization \gl{factor6} with $M=N$, from the definition of $Q_N^+$, from the statements (2) and (3) and from Remark~1. The statement (4) is valid in view of \gl{prod7} and smoothness (absence of pole(s)) for all elements of all matrix-valued coefficients of the operator ${\cal P}_N(H_+)$. 

To prove the statement (5) it is sufficient in view of \gl{prod7} to prove in the case, where all elements of all matrix-valued functions $V_+(x)$, $X_0^-(x)$, \dots, $X_{N-1}^-(x)$ and of the matrix $X_N^-$ are real-valued, that for any zero of the polynomial ${\cal P}_N(\lambda)$ with nonzero imaginary part there is the complex conjugate zero of this polynomial with the same multiplicity. The latter follows from the fact that for any chain of associated vector-functions of $H_+$ belonging to $\ker Q_N^-$ the chain of vector-functions with the complex conjugate components belongs to $\ker Q_N^-$ as well due to reality of all elements of all matrix-valued functions $X_0^-(x)$, \dots, $X_{N-1}^-(x)$ and of the matrix $X_N^-$ and is a chain of associated vector-functions of $H_+$ for the complex conjugate spectral value due to reality of all elements of $V_+(x)$.

The statement (7) is evident in view of \gl{prod7} and independence ${\cal P}_N(H_+)$ of orders of numberings mentioned in this statement. The final statement (9) follows from Theorem~1 and from the fact that one can choose a canonical basis in the kernel of the intertwining operator ${\cal P}_N(H_+)$ (it intertwines $H_+$ with $H_+$) such that (i) this basis contains the vector-functions $\Phi_l^-(x)$, $l=1$, \dots, $nN$ and (ii) for any eigenvalue of $T^+$ there are $2n$ chains of equal lengths (equal to the multiplicity of the corresponding zero of ${\cal P}_N(\lambda)$) of formal associated vector-functions of $H_+$ for this eigenvalue in this basis. Theorem 2 is proved.\\

{\bf Corollary 1.} In the conditions of Theorem 2 with the help of the super-Hamiltonian \[{\bf H}=\begin{pmatrix}H_+&0\\0&H_-\end{pmatrix}\] and the nilpotent supercharges \[{\bf Q}=\begin{pmatrix}0&Q_N^+\\0&0\end{pmatrix},\quad{\bf \bar Q}=\begin{pmatrix}0&0\\Q_N^-&0\end{pmatrix},
\qquad{\bf Q}^2={\bf \bar Q}^2=0\] one can construct the following polynomial algebra of supersymmetry:
\begin{equation}\{{\bf Q},{\bf \bar Q}\}={\cal P}_N({\bf H}),\qquad[{\bf H},{\bf Q}]=[{\bf H},{\bf \bar Q}]=0.\la{supalg7}\end{equation}

\vskip1pc

{\bf Corollary 2.} In the conditions of Theorem 2 the following relations hold, \begin{equation}\det(\lambda I_{nN}-T^+)=\det(\lambda I_{nN}-T^-)={\cal P}_N^n(\lambda),\qquad\forall\lambda\in\Bbb C,\la{det7}\end{equation} where $T^-$ is the matrix $T$ for the intertwining operator $Q_N^+$ (cf. \gl{supalg7} and \gl{det7} with (17) in \cite{anso03} or (43) in \cite{ancaso07}).

\section{Minimization of an intertwining operator and criterion of minimizability\la{secmin}}

It is evident that if we multiply a matrix $n\times n$ operator $Q_N^-$ intertwining a matrix $n\times n$ Hamiltonians $H_+$ and $H_-$ in accordance with \gl{splet} by a polynomial of the Hamiltonian $H_+$ from the right as follows, \[Q_N^-\Big[\sum_{l=0}^LA_l^+H_+^l\Big],\] where $A_l^+$, $l=0$, \dots, $L$ are constant matrices $n\times n$ commuting with $H_+$, or by a polynomial of the Hamiltonian $H_-$ from the left as follows, \[\Big[\sum_{l=0}^{L'}A_l^-H_-^l\Big]Q_N^-,\] where $A_l^-$, $l=0$, \dots, $L'$ are constant matrices $n\times n$ commuting with $H_-$, then such products are again a matrix $n\times n$ intertwining operators for the same Hamiltonians $H_+$ and $H_-$:
\[\Big\{Q_N^-\Big[\sum_{l=0}^LA_l^+H_+^l\Big]\Big\}H_+=Q_N^-H_+\Big[\sum_{l=0}^LA_l^+H_+^l\Big]=H_-\Big\{Q_N^-\Big[\sum_{l=0}^LA_l^+H_+^l\Big]\Big\},\]
\[\Big\{\Big[\sum_{l=0}^{L'}A_l^-H_-^l\Big]Q_N^-\Big\}H_+=\Big[\sum_{l=0}^{L'}A_l^-H_-^l\Big]H_-Q_N^-=H_-\Big\{\Big[\sum_{l=0}^{L'}A_l^-H_-^l\Big]Q_N^-\Big\}.\]
Thus, there is the problem of simplification of a matrix intertwining operator by separation from its right-hand side or from its left-hand side or both a superfluous polynomial in the corresponding Hamiltonian factor(s).

This section contains three definitions of minimizability in different senses of a matrix intertwining operator and the criterion of minimizability in the first of these senses of such operator.\\

{\bf Definition 1.} We define that a matrix $n\times n$ linear differential operator of $N$-th order $Q_N^-$ that intertwines a matrix $n\times n$ Hamiltonians $H_+$ and $H_-$ in accordance with \gl{splet} is {\it weakly minimizable} if this operator can be represented in the following form,
\[Q_N^-=P_M^-\Big[\sum_{l=0}^La_lH_+^l\Big],\]
where $a_l\in\Bbb C$, $l=1$, \dots, $L$, $a_L\ne0$, $1\leqslant L\leqslant N/2$ and $P_M^-$ is a matrix $n\times n$ linear differential operator of $M$-th order, $M=N-2L$ that intertwines the Hamiltonians $H_+$ and $H_-$, so that $P_M^-H_+=H_-P_M^-$. Otherwise, the operator $Q_N^-$ is called by us {\it weakly non-nminimizable}.\\

{\bf Definition 2.} We define that a matrix $n\times n$ linear differential operator of $N$-th order $Q_N^-$ that intertwines a matrix $n\times n$ Hamiltonians $H_+$ and $H_-$ in accordance with \gl{splet} is {\it strongly minimizable from the right} if this operator can be represented in the following form,
\[Q_N^-=P_M^-\Big[\sum_{l=0}^LA^+_lH_+^l\Big],\]
where $A^+_l$ is a constant matrix $n\times n$ that commutes with the Hamiltonian $H_+$, $l=1$, \dots, $L$, $A^+_L\ne0$, $1\leqslant L\leqslant N/2$ and $P_M^-$ is a matrix $n\times n$ linear differential operator of $M$-th order, $M=N-2L$ that intertwines the Hamiltonians $H_+$ and $H_-$, so that $P_M^-H_+=H_-P_M^-$. Otherwise, the operator $Q_N^-$ is called by us {\it strongly non-minimizable from the right}.\\

{\bf Definition 3.} We define that a matrix $n\times n$ linear differential operator of $N$-th order $Q_N^-$ that intertwines a matrix $n\times n$ Hamiltonians $H_+$ and $H_-$ in accordance with \gl{splet} is {\it strongly minimizable from the left} if this operator can be represented in the following form,
\[Q_N^-=\Big[\sum_{l=0}^LA^-_lH_-^l\Big]P_M^-,\]
where $A^-_l$ is a constant matrix $n\times n$ that commutes with the Hamiltonian $H_-$, $l=1$, \dots, $L$, $A^-_L\ne0$, $1\leqslant L\leqslant N/2$ and $P_M^-$ is a matrix $n\times n$ linear differential operator of $M$-th order, $M=N-2L$ that intertwines the Hamiltonians $H_+$ and $H_-$, so that $P_M^-H_+=H_-P_M^-$. Otherwise, the operator $Q_N^-$ is called by us {\it strongly non-nminimizable from the left}.\\

It is obvious that weak minimizability of a matrix intertwining operator is a partial case of left and right strong minimizabilities.\\ 

\noindent{\bf Theorem 3} (criterion of weak minimizability of a matrix intertwining operator).\linebreak {\it A~matrix $n\times n$ linear differential operator $Q_N^-$ of the $N$-th order with a constant nondegenerate matrix coefficient at $\partial^N$ that intertwines a matrix $n\times n$ Hamiltonians of Schr\"odinger form $H_+$ and $H_-$ in accordance with {\rm\gl{splet}} can be represented in the form
\begin{equation}Q_N^-=P_M^-\prod_{l=1}^s(\lambda_lI_n-H_+)^{\delta k_l},\la{minim8}\end{equation}
where
\renewcommand{\labelenumi}{\rm{(\theenumi)}}
\begin{enumerate}
\item $s\in\{0\}\cup\Bbb N;$
\item $\lambda_l\in\Bbb C$, $\l=1$, \dots, $s$ and $\lambda_l\ne \lambda_{l'}\Leftrightarrow l\ne l';$
\item $\delta k_l\in\Bbb N$, $l=1$, \dots, $s;$
\item $P_M^-$ is a weakly non-minimizable matrix $n\times n$ linear differential operator of the $M$-th order that intertwines the Hamiltonians $H_+$ and $H_-$, so that $P_M^-H_+=H_-P_M^-$, 
\end{enumerate}
if and only if
\renewcommand{\labelenumi}{\rm{(\theenumi)}}
\begin{enumerate}
\item all numbers $\lambda_l$, $l=1$, \dots, $s$ belong to the spectrum of the matrix $T^+$, {\it i.e.} the matrix $T$ of $Q_N^-$,
and there are no equal numbers in the set $\lambda_l$, $l=1$, \dots, {$s$}$;$
\item there are $2n$ Jordan cells in a normal form of the matrix $T^+$ for any its eigenvalue from the set $\lambda_l$, $l=1$, \dots, $s;$
\item there are no $2n$ Jordan cells in a normal form of the matrix $T^+$ for any its eigenvalue that does not belong to the set $\lambda_l$, $l=1$, \dots, $s;$
\item $\delta k_l$ is the minimal of the orders of Jordan cells corresponding to the eigenvalue $\lambda_l$ in a normal form of the matrix $T^+$, $l=1$, \dots, $s.$
\end{enumerate}}

\vskip1pc

{\bf Remark 3.} A normal (Jordan) form of the matrix $T$ of a matrix $n\times n$ linear differential intertwining operator $Q_N^-$ of arbitrary order $N$ that intertwines a matrix $n\times n$ Hamiltonians of Schr\"odinger form $H_+$ and $H_-$ in accordance with {\rm\gl{splet}} cannot have more than $2n$ Jordan blocks for the same eigenvalue $\lambda$, since otherwise $\ker(\lambda I_n-H_+)$ contains more than $2n$ linearly independent vector-functions.\\

{\bf Proof.} The statement of Theorem 3 is valid by virtue of Theorem 1 and the following facts:
\renewcommand{\labelenumi}{\rm{(\theenumi)}}
\begin{enumerate}
\item any operator of the form $(\lambda I_n-H_+)^{\delta k}$, $\lambda\in\Bbb C$, $\delta k\in\Bbb N$ intertwines the Hamiltonian $H_+$ with himself;
\item any canonical basis in the kernel of any operator of the form $(\lambda I_n-H_+)^{\delta k}$, $\lambda\in\Bbb C$, $\delta k\in\Bbb N$ consists of $2n$ chains of equal lengths equal to $\delta k$ of formal associated vector-functions of the Hamiltonian $H_+$ for the spectral value $\lambda$;
\item a matrix $n\times n$ linear differential operator with nondegenerate matrix coefficient at $\partial$ in the highest degree is uniquely defined by its kernel;
\item the Wronskian of all elements of any basis in the kernel of a matrix linear differential operator with constant nondegenerate matrix coefficient at $\partial$ in the highest degree and with sufficiently smooth other matrix-valued coefficients defined on entire real axis does not vanish on the real axis.
\end{enumerate}
Theorem 3 is proved.\\

{\bf Remark 4.} In the conditions of Theorem 3 the matrix $n\times n$ coefficient of $P_M^-$ at $\partial$ in the highest degree is equal obviously to $X_N^-$ (see \gl{splet}) and any element of any other matrix-valued coefficient of $P_M^-$ is smooth (without pole(s)) in view of \gl{minim8} and smoothness of all elements of all matrix-valued coefficients of $Q_N^-$.\\

{\bf Corollary 3.} In the conditions of Theorem 3 the following relations hold,
\[M=N-2\sum_{l=1}^s\delta k_l={1\over n}\sum_{l=1}^tk_l,\]
where 

$t$ is the total number of different eigenvalues of the matrix $T^+$;

$k_l$ is the difference of the algebraic multiplicity of $\lambda_l$ in the spectrum of $T^+$ and $n\delta k_l$, $l=1$, \dots, $s$;

$\lambda_l$, $l=s+1$, \dots, $t$ are eigenvalues of the matrix $T^+$ such that there are no equal between them and there are no $2n$ Jordan blocks in a normal (Jordan) form of the matrix $T^+$ for any of these eigenvalues;

$k_l$ is the algebraic multiplicity of $\lambda_l$ in the spectrum of $T^+$, $l=s+1$, \dots, $t$.

\section{Reducibility and irreducibility of an intertwining operator\la{secred}}
\subsection{Definitions of different types of reducibility and irreducibility for matrix intertwining operators}

Let us introduce the notions of regularly reducible and irregucible,  singularly reducible and irreducible and regularly absolutely irreducible matrix intertwining operators.\\

{\bf Definition 4.} We define that a matrix $n\times n$ linear differential operator $Q_N^-$ of the $N$-th order that intertwines a matrix $n\times n$ Hamiltonians $H_+$ and $H_-$ of Schr\"odinger form in accordance with \gl{splet} is {\it regularly reducible} if there are a matrix $n\times n$  linear differential intertwining operators $K_{N-M}^-$ and $P_M^-$ of the orders $N-M$ and $M$, $0<M<N$ respectively and a matrix $n\times n$ intermediate Hamiltonian of Schr\"odinger form $H_M\equiv -I_n\partial^2+V_M(x)$ such that the relations \begin{equation}Q_N^-=K_{N-M}^-P_M^-,\qquad K^-_{N-M}H_M=H_-K^-_{N-M},\qquad
P_M^-H_+=H_MP_M^-\la{interm2.4}\end{equation}
hold and all elements of the potential $V_M(x)$ and of all matrix-valued coefficients of the operators $K_{N-M}^-$ and $P_M^-$ are smooth. Otherwise the operator $Q_N^-$ is called by us {\it regularly irreducible}.\\

{\bf Definition 5.} We define that a matrix $n\times n$ linear differential operator $Q_N^-$ of the $N$-th order that intertwines a matrix $n\times n$ Hamiltonians $H_+$ and $H_-$ of Schr\"odinger form in accordance with \gl{splet} is {\it singularly reducible} if there are a matrix $n\times n$  linear differential intertwining operators $K_L^-$ and $P_M^-$ of the orders $L$ and $M$ respectively, $L+M>N$ and a matrix $n\times n$ intermediate Hamiltonian of Schr\"odinger form $H_{L,M}\equiv -I_n\partial^2+V_{L,M}(x)$ such that the relations \begin{equation}Q_N^-=K_L^-P_M^-,\qquad K^-_LH_{L,M}=H_-K^-_L,\qquad
P_M^-H_+=H_{L,M}P_M^-\la{interm2.20}\end{equation}
hold and all elements of the potential $V_{L,M}(x)$ and of all matrix-valued coefficients of the operators $K_L^-$ and $P_M^-$ are smooth. Otherwise the operator $Q_N^-$ is called by us {\it singularly irreducible}.\\

{\bf Remark 5.} It is obvious that both matrix coefficients of $K_L^-$ and $P_M^-$ at $\partial$ in the eldest powers in conditions of Definition 5 are degenerate matrices and their product is zero matrix.\\

{\bf Remark 6.} It is evident that one can define more special types of reducibility, for example, for the cases where both Hamiltonians $H_+$ and $H_-$ are Hermitian or symmetric with respect to transposition or possess by a potentials all elements of which are real-valued. It is natural to require for the intermediate Hamiltonian $H_M$ in these cases to be respectively as well Hermitian or symmetric with respect to transposition or possess by a potential all elements of which are real-valued.\\

{\bf Definition 6.} We define that a matrix $n\times n$ linear differential operator $Q_N^-$ of the $N$-th order that intertwines a matrix $n\times n$ Hamiltonians $H_+$ and $H_-$ of Schr\"odinger form in accordance with \gl{splet} is {\it regularly absolutely irreducible} if for any $M$, $0<M<N$ there are no a matrix $n\times n$ linear differential intertwining operators $K_{N-M}^-$ and $P_M^-$ of the orders $N-M$ and $M$ respectively and a matrix $n\times n$ intermediate Hamiltonian $H_M\equiv -I_n\partial^2+V_M(x)$, even with the potential $V_M(x)$ and the matrix-valued coefficients of the operators $K_{N-M}^-$ and $P_M^-$ element(s) of which possess by a pole singularity(-ies), such that the relations \gl{interm2.4} hold.\\

Example of 2-nd order matrix $2\times2$ intertwining operator that is regularly absolutely irreducible but is reducible singularly was presented by the author at PHHQP XVII (see abstract in \cite{SokBadHonnef}).

\subsection{Criterion of regular reducibility of a matrix intertwining operator}

The following theorem contains the sufficient condition of reducibility of a matrix intertwining operator $Q_N^-$.\\

\noindent{\bf Theorem 4} (sufficient condition of regular reducibility of a matrix intertwining operator).\linebreak {\it Suppose that
\renewcommand{\labelenumi}{\rm{(\theenumi)}}
\begin{enumerate}
\item a matrix $n\times n$ Hamiltonians $H_+$ and $H_-$ of the form {\rm\gl{h+h-2.1}} are intertwined by a matrix $n\times n$ linear differential operator of the $N$-th order $Q_N^-$ of the form {\rm\gl{splet}} with constant nondegenerate matrix coefficient $X_N^-$ at $\partial^N$ in accordance with {\rm\gl{splet};}
\item a vector-functions $\Phi_l^-(x)\equiv(\varphi_{l1}^-(x),\ldots,\varphi_{ln}^-(x))^t$, $l=1$, \dots, $nM$, $1\leqslant M<N$ belong to $\ker Q_N^-$ and can be divided into a chains of formal associated vector-functions of the Hamiltonian $H_+;$
\item the Wronskian
\begin{equation}\begin{vmatrix}\varphi^-_{11}&\dots&\varphi^-_{1n}& \varphi^{-\prime}_{11}&\dots&\varphi^{-\prime}_{1n}&\ldots&(\varphi^-_{11})^{(M-1)}&\dots&(\varphi^-_{1n})^{(M-1)}\\
\varphi^-_{21}&\dots&\varphi^-_{2n}&\varphi^{-\prime}_{21}&\dots&\varphi^{-\prime}_{2n}&\ldots&
(\varphi^-_{21})^{(M-1)}&\dots&(\varphi^-_{2n})^{(M-1)}\\
\vdots&\ddots&\vdots&\vdots&\ddots&\vdots&\ddots&\vdots&\ddots&\vdots\\
\varphi^-_{nM,1}\!\!&\dots&\!\!\varphi^-_{nM,n}&\varphi^{-\prime}_{nj,1}\!\!&\dots&\!\!\varphi^{-\prime}_{nM,n}\!\!&\ldots&\!\!(\varphi^-_{nM,1})^{(M-1)}\!\!&\dots&\!\!(\varphi^-_{nM,n})^{(M-1)}\end{vmatrix}\la{wron9.2}\end{equation} does not vanish on the real axis.
\end{enumerate}
Then there exist the matrix $n\times n$ linear differential operators $K_{N-M}^-$ and $P_M^-$  of the orders $N-M$ and $M$ respectively and the matrix $n\times n$ Hamiltonian of Schr\"odinger form $H_M\equiv -I_n\partial^2+V_M(x)$ such that$:$
\renewcommand{\labelenumi}{\rm{(\theenumi)}}
\begin{enumerate}
\item the matrix coefficients of $K_{N-M}^-$ and $P_M^-$ at $\partial^{N-M}$ and $\partial^M$ are equal to $X_N^-$ and~$I_n$ respectively and all elements of all other matrix-valued coefficients of $K_{N-M}^-$ and $P_M^-$ and of the potential $V_M(x)$ are smooth $($without pole$($s$));$
\item  the relations {\rm\gl{interm2.4}} hold$;$
\item the vector-functions $\Phi_l^-(x)$, $l=1$, \dots, $nM$ form a canonical basis in $\ker P_M^-,$
\end{enumerate}
and the operator $Q_N^-$ is reducible.}\\

{\bf Proof.} The matrix $n\times n$ linear differential operator of the $M$-th order $P_M^-$ and the matrix $n\times n$ Hamiltonian of Schr\"odinger form $H_M$, that satisfy the statements (1)~-- (3) in the part that applies to them, can be constructed by the method of the Sec. 4 of \cite{sokolov15}. The facts that the operator $P_M^-$ can be separated from the right-hand side of the operator $Q_N^-$ and that the operator $K_{N-M}^-$, being the rest of $Q_N^-$ under such separation, intertwines the Hamiltonians $H_M$ and $H_-$ in accordance with \gl{interm2.4} can be proved in the same way as in the proof of Theorem 1. The statements about properties of coefficients of the operator $K_{N-M}^-$ and about its order are evident in view of the factorization $Q_N^-=K_{N-M}^-P_M^-$. Thus, the intertwining operator $Q_N^-$ is regularly reducible. Theorem 4 is proved.\\

{\bf Remark 7.} Theorem 4 is a criterion in fact of regular reducibility of a matrix intertwining operator $Q_N^-$ since if this operator is regularly reducible then any canonical basis in the kernel of the intertwining operator $P_M^-$ from Definition 4 satisfies the conditions (2) and (3) of Theorem~4.

\subsection{Existence of regularly absolutely irreducible intertwining operators of any order}

Unlike to the scalar case $n=1$ where regularly absolutely irreducible intertwining operators absent (see Lemma 1 in \cite{anso03}) we have in the matrix case with any $n\geqslant2$ regularly absolutely irreducible intertwining operators of any order $N\geqslant2$. Existence of such operators is provided by the proposed below method of constructing of matrix $n\times n$ absolutely irreducible intertwining operators of any order $N$ of some type and by the fact that for any $m\in{\Bbb N}$ there are a scalar Hamiltonians any of which possesses by a chain of $m$ formal associated functions with the Wronskian that does not vanish on the real axis (the examples of such Hamiltonians can be found in \cite{anso11}).

Let us start with the case $n=2$. Consider two scalar Hamiltonians of Schr\"odinger type
\[h_1=-\partial^2+v_1(x),\qquad h_2=-\partial^2+v_2(x),\] where $v_1(x)$ and $v_2(x)$ are complex-valued, in general, and sufficiently smooth functions defined on the entire axis, and two chains of a formal associated functions of these Hamiltonians for a spectral value $\lambda_0\in\Bbb C$:
\begin{eqnarray}h_1\varphi_{1,0}=\lambda_0\varphi_{1,0},&\qquad&(h_1-\lambda_0)\varphi_{1,l}=\varphi_{1,l-1},\quad l=1,\dots,2N-1,\nonumber\\
h_2\varphi_{2,0}=\lambda_0\varphi_{2,0},&\qquad&(h_2-\lambda_0)\varphi_{2,l}=\varphi_{2,l-1},\quad l=1,\dots,N-1\nonumber\end{eqnarray}
such that the Wronskians \[W_{1,N}(x)\!\equiv\!\begin{vmatrix}\varphi_{1,0}&\varphi'_{1,0}&\dots&\varphi^{(N-1)}_{1,0}\\
\varphi_{1,1}&\varphi'_{1,1}&\dots&\varphi^{(N-1)}_{1,1}\\
\vdots&\vdots&\ddots&\vdots\\
\varphi_{1,N-1}\!\!&\!\!\varphi'_{1,N-1}\!\!&\dots&\!\!\varphi^{(N-1)}_{1,N-1}\end{vmatrix},\qquad
W_{2,N}(x)\!\equiv\!\begin{vmatrix}\varphi_{2,0}&\varphi'_{2,0}&\dots&\varphi^{(N-1)}_{2,0}\\
\varphi_{2,1}&\varphi'_{2,1}&\dots&\varphi^{(N-1)}_{2,1}\\
\vdots&\vdots&\ddots&\vdots\\
\varphi_{2,N-1}\!\!&\!\!\varphi'_{2,N-1}\!\!&\dots&\!\!\varphi^{(N-1)}_{2,N-1}\end{vmatrix}\] do not vanish on the entire axis. Then the vector-functions \[\Phi_l^-(x)\!\equiv\!\begin{pmatrix}\varphi_{1,l}(x)\\0\end{pmatrix},\quad l\!=\!0,\ldots,N\!-\!1,\qquad
\Phi_l^-(x)\!\equiv\!\begin{pmatrix}\varphi_{1,l}(x)\\\varphi_{2,l-N}(x)\end{pmatrix},\quad l\!=\!N,\ldots,2N\!-\!1\] form obviously the chain of formal associated vector-functions of the matrix $2\times2$ Hamiltonian of Schr\"odinger type \[H_+=\begin{pmatrix}h_1&0\\0&h_2\end{pmatrix}\] for the spectral value $\lambda_0$,
\[H_+\Phi_0^-=\lambda_0\Phi_0^-,\qquad (H_+-\lambda_0I_n)\Phi_l^-=\Phi_{l-1}^-,\qquad l=1,\ldots,2N-1\]  and one can check with the help of columns permutations that the following equalities for the Wronskian of $\Phi_l^-(x)$, $l=0$, \dots, $2N-1$ hold, 
\begin{equation}\begin{vmatrix}\varphi_{1,0}&0&\varphi'_{1,0}&0&\ldots&\varphi^{(N-1)}_{1,0}&0\\
\varphi_{1,1}&0&\varphi'_{1,1}&0&\ldots&\varphi^{(N-1)}_{1,1}&0\\
\vdots&\vdots&\vdots&\vdots&\ddots&\vdots&\vdots\\
\varphi_{1,N-1}&0&\varphi'_{1,N-1}&0&\ldots&\varphi^{(N-1)}_{1,N-1}&0\\
\varphi_{1,N}&\varphi_{2,0}&\varphi'_{1,N}&\varphi'_{2,0}&\ldots&\varphi^{(N-1)}_{1,N}&\varphi^{(N-1)}_{2,0}\\
\varphi_{1,N+1}&\varphi_{2,1}&\varphi'_{1,N+1}&\varphi'_{2,1}&\ldots&\varphi^{(N-1)}_{1,N+1}&\varphi^{(N-1)}_{2,1}\\
\vdots&\vdots&\vdots&\vdots&\ddots&\vdots&\vdots\\
\varphi_{1,2N-1}&\varphi_{2,N-1}&\varphi'_{1,2N-1}&\varphi'_{2,N-1}&\ldots&\varphi^{(N-1)}_{1,2N-1}&\varphi^{(N-1)}_{2,N-1}
\end{vmatrix}\qquad\qquad\qquad\qquad\quad\la{wron9.3}\end{equation}
\begin{eqnarray}&=&(-1)^{N(N-1)/2}\begin{vmatrix}\varphi_{1,0}&\varphi'_{1,0}&\ldots&\varphi^{(N-1)}_{1,0}&0&0&\ldots&0\\
\varphi_{1,1}&\varphi'_{1,1}&\ldots&\varphi^{(N-1)}_{1,1}&0&0&\ldots&0\\
\vdots&\vdots&\ddots&\vdots&\vdots&\vdots&\ddots&\vdots\\
\varphi_{1,N-1}&\varphi'_{1,N-1}&\ldots&\varphi^{(N-1)}_{1,N-1}&0&0&\ldots&0\\
\varphi_{1,N}&\varphi'_{1,N}&\ldots&\varphi^{(N-1)}_{1,N}&\varphi_{20}&\varphi'_{2,0}&\ldots&\varphi^{(N-1)}_{2,0}\\
\varphi_{1,N+1}&\varphi'_{1,N+1}&\ldots&\varphi^{(N-1)}_{1,N+1}&\varphi_{2,1}&\varphi'_{2,1}&\ldots&\varphi^{(N-1)}_{2,1}\\
\vdots&\vdots&\ddots&\vdots&\vdots&\vdots&\ddots&\vdots\\
\varphi_{1,2N-1}&\varphi'_{1,2N-1}&\ldots&\varphi^{(N-1)}_{1,2N-1}&\varphi_{2,N-1}&\varphi'_{2,N-1}&\ldots&\varphi^{(N-1)}_{2,N-1}
\end{vmatrix}\nonumber\\
\nonumber\\
&=&(-1)^{N(N-1)/2}W_{1,N}(x)W_{2,N}(x).\nonumber\end{eqnarray}
Thus, the Wronskian of $\Phi_l^-(x)$, $l=0$, \dots, $2N-1$ does not vanish on the real axis and by virtue of the results of the Sec. 4 from \cite{sokolov15} there are the matrix $2\times 2$ Hamiltonian of Schr\"odinger form $H_-$ and the matrix $2\times2$ linear differential operator $Q_N^-$ of the order~$N$ such that:
\renewcommand{\labelenumi}{\rm{(\theenumi)}}
\begin{enumerate}
\item the matrix coefficient of $Q_N^-$ at $\partial^N$ is equal to arbitrary nondegenerate matrix $X_N^-$ of the second order and all elements of all other matrix-valued coefficients of $Q_N^-$ and of the potential of the Hamiltonian $H_-$ are smooth;
\item the Hamiltonians $H_+$ and $H_-$ are intertwined by the operator $Q_N^-$ in accordance with \gl{splet};
\item the vector-functions $\Phi_l^-(x)$, $l=0$, \dots, $2N-1$ form a canonical basis in $\ker Q_N^-$.
\end{enumerate}
In this case, regular absolute irreducibility of the intertwining operator $Q_N^-$ takes place in view of the following contradiction:
\renewcommand{\labelenumi}{\rm{(\theenumi)}}
\begin{enumerate}
\item if there are a matrix $2\times2$ linear differential intertwining operators $K_{N-M}^-$ and $P_M^-$ of the orders $N-M$ and $M$ respectively, $1\leqslant M\leqslant N-1$ and a matrix $2\times2$ intermediate Hamiltonian $H_M$ of Schr\"odinger form such that the relations \gl{interm2.4} hold, then obviously the constant matrix coefficient of $P_M^-$ at $\partial^M$ is nondegenerate and the Wronskian of elements of any basis in $\ker P_M^-$ is different from identical zero on the real axis;
\item any canonical basis in $\ker P_M^-$ consists either of the vector-functions $\Phi_l^-(x)$, $l=0$, \dots, $2M-1$ or of their linear combinations, but the Wronskian of the vector-functions $\Phi_l^-(x)$, $l=0$, \dots, $2M-1$ is identical zero on the real axis since it coincides with the minor of the $2M$-th order of the Wronskian \gl{wron9.3} that is situated in the upper left corner of \gl{wron9.3} and this minor contains either zero column(s) (if $1\leqslant M\leqslant N/2$) or linearly dependent for any fixed $x$ columns 
\[\begin{pmatrix}0\\\vdots\\0\\\varphi_{2,0}(x)\\\vdots\\\varphi_{2,2M-N-1}(x)\end{pmatrix},\quad
\begin{pmatrix}0\\\vdots\\0\\\varphi'_{2,0}(x)\\\vdots\\\varphi'_{2,2M-N-1}(x)\end{pmatrix},\quad\ldots,\quad
\begin{pmatrix}0\\\vdots\\0\\\varphi^{(M-1)}_{2,0}(x)\\\vdots\\\varphi^{(M-1)}_{2,2M-N-1}(x)\end{pmatrix}\]
(if $N/2<M\leqslant N-1$).
\end{enumerate}

A generalization of the described method to the case $n>2$ is straightforward. Consider a scalar Hamiltonians of Schr\"odinger form \[h_l=-\partial^2+v_l(x),\qquad l=1,\ldots,n,\]
where the potentials $v_l(x)$, $l=1$, \dots, $n$ are complex-valued, in general, and sufficiently smooth functions defined on the entire axis, and a chains of formal associated functions of these Hamiltonians for a spectral value $\lambda_0\in\Bbb C$ such that 
\[h_l\varphi_{l,0}\!=\!\lambda_0\varphi_{l,0},\qquad (h_l\!-\!\lambda_0)\varphi_{l,j}\!=\!\varphi_{l,j-1},\quad j\!=\!1,\ldots,N(n\!-\!l\!+\!1)\!-\!1,\qquad l\!=\!1,\ldots,n\]
and the Wronskians
\[W_{l,N}(x)\equiv\begin{vmatrix}\varphi_{l,0}(x)&\varphi'_{l,0}(x)&\ldots&\varphi^{(N-1)}_{l,0}(x)\\
\varphi_{l,1}(x)&\varphi'_{l,1}(x)&\ldots&\varphi^{(N-1)}_{l,1}(x)\\
\vdots&\vdots&\ddots&\vdots\\
\varphi_{l,N-1}(x)&\varphi'_{l,N-1}(x)&\ldots&\varphi^{(N-1)}_{l,N-1}(x)
\end{vmatrix},\qquad l=1,\ldots,n\]
do not vanish on the real axis. Then the vector-functions 
\[\Phi_l^-(x)\equiv\begin{pmatrix}\varphi_{1,l}(x)\\\varphi_{2,l-N}(x)\\\varphi_{3,l-2N}(x)\\\vdots\\\varphi_{n,l-(n-1)N}(x)\end{pmatrix},
\qquad l=0,\ldots,nN-1,\qquad\qquad\qquad\qquad\qquad\qquad\]
\[\qquad\qquad\qquad\qquad\qquad\qquad\qquad\qquad\varphi_{l.j}(x)\equiv0,\qquad j<0,\quad l=0,\ldots,nN-1\] form obviously the chain of formal associated vector-functions of the matrix $n\times n$ Hamiltonian of Schr\"odinger type
\[H_+={\rm {diag}}\,(h_1,h_2,\ldots,h_n)\]
for the spectral value $\lambda_0$,
\[H_+\Phi_0^-=\lambda_0\Phi_0^-,\qquad (H_+-\lambda_0I_n)\Phi_l^-=\Phi_{l-1}^-,\qquad l=1,\ldots,nN-1\] 
and one can check with the help of columns permutations that the following equality for the Wronskian of $\Phi_l^-(x)$, $l=0$, \dots, $nN-1$ holds, 
\[\begin{vmatrix}\varphi_{1,0}&\ldots&\!\!\varphi_{n,-(n-1)N}\!\!&\varphi'_{1,0}&\ldots&\!\!\varphi'_{n,-(n-1)N}\!\!&\ldots&\varphi^{(N-1)}_{1,0}&\ldots&\!\!\varphi^{(N-1)}_{n,-(n-1)N}\\
\varphi_{1,1}&\ldots&\!\!\varphi_{n,1-(n-1)N}\!\!&\varphi'_{1,1}&\ldots&\!\!\varphi'_{n,1-(n-1)N}\!\!&\ldots&\varphi^{(N-1)}_{1,1}&
\ldots&\!\!\varphi^{(N-1)}_{n,1-(n-1)N}\\
\vdots&\ddots&\vdots&\vdots&\ddots&\vdots&\ddots&\vdots&\ddots&\vdots\\
\varphi_{1,nN-1}\!\!&\ldots&\varphi_{n,N-1}&\!\!\varphi'_{1,nN-1}\!\!&\ldots&\varphi'_{n,N-1}&\ldots&\!\!\varphi^{(N-1)}_{1,nN-1}\!\!&
\ldots&\varphi^{(N-1)}_{n,N-1}\end{vmatrix}\]

\[=(-1)^{n(n-1)N(N-1)/4}W_{1,N}(x)W_{2,N}(x)\cdot\ldots\cdot W_{n,N}(x).\]
Thus, the Wronskian of $\Phi_l^-(x)$, $l=0$, \dots, $nN-1$ does not vanish on the entire axis. Constructing of the matrix $n\times n$ 
Hamiltonian $H_-$ of Schr\"odinger form and the matrix $n\times n$ linear differential intertwining operator $Q_N^-$ of the order $N$ that possess by the properties analogous to their properties (1) -- (3) in the case $n=2$ and proving of absolute irreducibility of the operator $Q_N^-$ can be produced in the same way as in the indicated case.

\section{Polynomial SUSY in general case\la{seccon}}

The following theorem establishes the fact that for any matrix $n\times n$ linear differential intertwining operator $Q_N^-$ there is the weakly non-minimizable matrix $n\times n$ linear differential operator $Q_{N'}^+$ that intertwines the same Hamiltonians in the opposite direction and such that the product $Q_{N'}^+Q_N^-$ is a polynomial of the Hamiltonian $H_+$. Moreover, this theorem contains the formula for calculation of $N'$, {\it i.e.} of the order of the operator~$Q_{N'}^+$.\\

\noindent{\bf Theorem 5} (on existence of ``conjugate'' intertwining operator $Q_{N'}^+$ for any intertwining operator $Q_N^-$ such that $Q_{N'}^+Q_N^-$ is polynomial of $H_+$). 

\noindent{\it Suppose that
\renewcommand{\labelenumi}{\rm{(\theenumi)}}
\begin{enumerate}
\item a matrix $n\times n$ Hamiltonians $H_+$ and $H_-$ of the form {\rm\gl{h+h-2.1}} are intertwined by a matrix $n\times n$ linear differential operator of the $N$-th order $Q_N^-$ of the form {\rm\gl{splet}} with constant nondegenerate matrix coefficient $X_N^-$ at $\partial^N$ in accordance with {\rm\gl{splet};}
\item $T^+$ is the matrix $T$ of the intertwining operator $Q_N^-;$
\item $\lambda_l$, $l=1$, \dots, $L$ is the set of all different eigenvalues of the matrix $T^+$ $($i.e. any eigenvalue of the matrix $T^+\!$ is contained in this set and there no equal numbers in~it$);$
\item $g_l^-$ is the geometric multiplicity of the eigenvalue $\lambda_l$ in the spectrum of the matrix $T^+$, $l=1$, \dots, $L;$
\item $k_{l,j}^-$, $j=1$, \dots, $g_l^-$ are the orders of Jordan blocks in a normal $($Jordan$)$ form of the matrix $T^+$ that correspond to the eigenvalue $\lambda_l$, $l=1$, \dots, $L;$ moreover, these blocks are numbered so that the sequence $k_{l,j}^-$, $j=1$, \dots, $g_l^-$ is nonincreasing, $l=1$, \dots, $L;$
\item the number $\varkappa_l$ is defined by the equality
\[\varkappa_l=\max_{1\leqslant j\leqslant g_l^-}k_{l,j}^-,\qquad l=1,\ldots,L;\]
\item $\mu_l$ is the number of Jordan blocks of the order $\varkappa_l$ corresponding to the eigenvalue $\lambda_l$ in a normal $($Jordan$)$
form of the matrix $T^+$, $l=1$, \dots, $L.$
\end{enumerate}
Then there is the weakly non-minimizable matrix $n\times n$ linear differential operator $Q_{N'}^+$ that intertwines the Hamiltonians $H_+$ and $H_-$ as follows, 
\begin{equation}H_+Q_{N'}^+=Q_{N'}^+H_-\la{int'10}\end{equation}
and such that$:$
\renewcommand{\labelenumi}{\rm{(\theenumi)}}
\begin{enumerate}
\item the matrix coefficient of the operator $Q_{N'}^+$ at $\partial^{N'}$ is equal to $(-1)^{\varkappa_1+\ldots+\varkappa_L}(X_N^-)^{-1}$ and all elements of all other matrix-valued coefficients of the operator $Q_{N'}^+$ are smooth$;$
\item all elements of all matrix-valued coefficients of the operator $Q_{N'}^+$ are complex-valued, in general, but are real-valued in the case where all elements of all matrix-valued coefficients of the operator $Q_N^-$ and of the potential of the Hamiltonian $H_+$ are real-valued$;$
\item the following equality for the product $Q_{N'}^+Q_N^-$ holds,
\begin{equation}Q_{N'}^+Q_N^-=\prod_{l=1}^L(H_+-\lambda_lI_n)^{\varkappa_l};\la{pol10}\end{equation}
\item the following representation for the order $N'$ of the operator $Q_{N'}^+$ takes place,
\[N'=-N+2\sum_{l=1}^L\varkappa_l;\]
\item any eigenvalue of the matrix $T^-$, i.e of the matrix $T$ of the intertwining operator $Q_{N'}^+$ belongs to the set $\lambda_l$, $l=1$, \dots, $L;$ 
\item the number $\lambda_l$ belongs to the spectrum of the matrix $T^-$ iff $\mu_l<2n$, where $l=1$, \dots, $L$ is arbitrarily fixed$;$
\item the geometric multiplicity $g_l^+$ of the number $\lambda_l$ in the spectrum of the matrix $T_-$ can be found with the help of the equality 
\[g_l^+=2n-\mu_l,\qquad l=1, \dots, L\] $($if $\lambda_l$ does not belong to the spectrum of the matrix $T^-$ then the geometric multiplicity of $\lambda_l$ in this spectrum is zero by definition, $l=1$, \dots, $L);$
\item if $\lambda_l$ belongs to the spectrum of the matrix $T^-$ then Jordan blocks corresponding to $\lambda_l$ in a normal $($Jordan$)$ form of $T^-$ can be numbered so that their orders $k_{l,j}^+$, $j=1$, \dots, $g_l^+$ form a nonincreasing sequence and the following representation holds for these orders, \[k^+_{l,j}=\begin{cases}\varkappa_l,&1\leqslant j\leqslant 2n-g_l^-,\\
\varkappa_l-k_{l,2n-j+1}^-,&2n-g_l^-+1\leqslant j\leqslant g_l^+,\end{cases}\qquad l=1,\dots,L;\]
\item if a vector-functions $\Phi_l^-(x)$, $l=1$, \dots, $2n(\varkappa_1+\ldots+\varkappa_L)$ form a canonical basis in the kernel of the polynomial of $H_+$ from the right-hand side of the {\rm\gl{pol10}} and the set of these vector-functions includes all elements of a canonical basis in $\ker Q_N^-$, then the nonzero vector-functions from the set  $Q_N^-\Phi_l^-(x)$, $l=1$, \dots, $2n(\varkappa_1+\ldots+\varkappa_L)$ after appropriate renumbering form a canonical basis in $\ker Q_{N'}^+;$
\item there is no a nonzero matrix $n\times n$ linear differential operator $P_M^+$ of the $M$-th order, $M<N'$ such that the intertwining $H_+P_M^+=P_M^+H_-$ takes place and the product $P_M^+Q_N^-$ is a polynomial of the Hamiltonian~$H_+$ coefficients of which are numbers $($not matrices$).$
\end{enumerate}}

\vskip1pc

{\bf Proof.} Existence of a matrix $n\times n$ linear differential operator $Q_{N'}^+$ of the $N'$-th order that intertwines the Hamiltonians $H_+$ and $H_-$ in accordance with \gl{int'10} and such that the statements (1), (3) and (4) are valid is evident in view of Theorem 4 and the results of the Sec. 4 from \cite{sokolov15} if to take as $Q_N^-$ and $P_M^-$ in the conditions of Theorem 4 the operators from the right-hand part of \gl{pol10} and $\tilde Q_N^-=(X_N^-)^{-1}Q_N^-$ respectively. The statement (9) follows from Theorem 1 and from the results of the Sec. 4 from \cite{sokolov15}. The statement (5) follows from the statement (9) in view of the Sec. 2.3 from \cite{sokolov15}. The statements (6) -- (8) take place by virtue of the statement (9) and the fact that any canonical basis in the kernel of the right-hand part of \gl{pol10} contains $2n$ chains of formal associated vector-functions of the Hamiltonian $H_+$ of equal lengths $\varkappa_l$ for any $\lambda_l$,
$l=1$, \dots, $L$ and there are no other elements in this basis. The fact that the operator $Q_{N'}^+$ is weakly non-minimizable follows from Theorem 3, from statements (5)  and (7) and from the fact that $\mu_l\geqslant1$, $l=1$, \dots,~$L$.

To prove the statement (10), assume the contrary and consider the polynomial that describes dependence of $P_M^+Q_N^-$ from the Hamiltonian $H_+$. Since $M<N'$, so there is $l_0$ such that $1\leqslant l_0\leqslant L$ and the multiplicity of $\lambda_{l_0}$ as zero of the considered polynomial is less than $\varkappa_{l_0}$ (see \gl{pol10}). But then a formal associated vector-function of $(\varkappa_{l_0}-1)$-th order of the Hamiltonian $H_+$ for the spectral value $\lambda_{l_0}$ from a canonical basis in $\ker Q_N^-$ cannot belong to $\ker P_M^+Q_N^-$. It follows from this contradiction that the statement (10) is valid.

To prove the statement (2), it is sufficient to consider the case where all elements of the potential of $H_+$ and of all matrix-valued coefficients of $Q_N^-$ are real-valued since the opposite case is trivial. All elements of the potential of the Hamiltonian $H_-$ in this case are real-valued as well by virtue of~\gl{vmp2.1}. Assume in the considered case that there is one at least matrix-valued coefficient of $Q_{N'}^+$ with an element that possess by nonzero imaginary part. Then in view of \gl{int'10}, \gl{pol10} and the statement (1) the operator $P_M^+=(Q_{N'}^+-Q_{N'}^{+*})/(2i)$ is a nonzero matrix $n\times n$ linear differential operator of the order $M$, $M<N'$ such that the intertwining $H_+P_M^+=P_M^+H_-$ holds and the product $P_M^+Q_N^-$ is a polynomial of the Hamiltonian $H_+$ that contradicts to the statement (10). Hence, the statement (2) takes place and Theorem 5 is proved.\\

{\bf Corollary 4.} The numbers $N$ and $N'$ in the conditions of Theorem 5 are of the same parity, {i.e.} these numbers are either  both odd or both even.\\

{\bf Corollary 5.} If the intertwining operator $Q_N^-$ in the conditions of Theorem 5 is weakly non-minimizable then by virtue of Theorems 3 and 5 any eigenvalue of the matrix $T^+$ is an eigenvalue of the matrix $T^-$ and, conversely, any eigenvalue of the matrix $T^-$ is an eigenvalue of the matrix $T^+$.\\

{\bf Corollary 6.} If an intertwining operator $Q_N^-$ satisfies the conditions of Theorem 2 and Theorem 5 then the intertwining operator $Q_N^+$ of Theorem 2 is identical to the intertwining operator $Q_{N'}^+$ of Theorem 5 and, thus, $N'=N$ in this case.\\

{\bf Corollary 7.} The following relations take place in the conditions of Theorem 5 in view of \gl{splet},
$$Q_N^-Q_{N'}^+Q_N^-=Q_N^-\Big[\prod_{l=1}^L(H_+-\lambda_lI_n)^{\varkappa_l}\Big]=\Big[\prod_{l=1}^L(H_--\lambda_lI_n)^{\varkappa_l} \Big]Q_N^-,$$
wherefrom it follows that 
$$Q_N^-Q_{N'}^+=\prod_{l=1}^L(H_--\lambda_lI_n)^{\varkappa_l}.$$
Thus, in the conditions of Theorem 5 with the help of the super-Hamiltonian \[{\bf H}=\begin{pmatrix}H_+&0\\0&H_-\end{pmatrix}\] and the nilpotent supercharges \[{\bf Q}=\begin{pmatrix}0&Q_{N'}^+\\0&0\end{pmatrix},\quad{\bf \bar Q}=\begin{pmatrix}0&0\\Q_N^-&0\end{pmatrix},
\qquad{\bf Q}^2={\bf \bar Q}^2=0\] one can construct the following polynomial algebra of supersymmetry:
\begin{equation}\{{\bf Q},{\bf \bar Q}\}={\cal P}_{(N+N')/2}({\bf H}),\qquad[{\bf H},{\bf Q}]=[{\bf H},{\bf \bar Q}]=0,\la{supalg10}\end{equation}
where the polynomial
$${\cal P}_{(N+N')/2}(\lambda)\equiv\prod_{l=1}^L(\lambda-\lambda_l)^{\varkappa_l}.$$

\vskip1pc

{\bf Corollary 8.} In the conditions of Theorem 5 the following relation holds, 
\begin{equation}\det(\lambda I_{nN}-T^+)\det(\lambda I_{nN'}-T^-)={\cal P}_{(N+N')/2}^{2n}(\lambda),\qquad\forall\lambda\in\Bbb C\la{det10}\end{equation} 
(cf. \gl{supalg10} and \gl{det10} with (17) in \cite{anso03} or (43) in \cite{ancaso07}).\\

{\bf Remark 8.} It is evident that $Q_{N'}^+$ is the only nonzero weakly non-minimizable matrix $n\times n$ differential operator that intertwines $H_+$ and $H_-$ in the direction opposite to one for $Q_N^-$ and such that its product wit $Q_N^-$ from the right-hand side is a polynomial of $H_+$ with number (not matrix) coefficients.\\

Let us introduce the notion of {\it complement} for a matrix intertwining operator. The operation of constructing of the complement has some properties close to ones of Hermitian conjugation and transposing (see below) and it seems to be perspective in investigation of supersymmetry with matrix Hamiltonians.\\

{\bf Definition 7.} The intertwining operator $Q_{N'}^+$ that exists in accordance with Theorem~5 for any matrix $n\times n$ linear differential operator $Q_N^-$ of the $N$-th order with nondegenerate matrix coefficient at $\partial^N$, which intertwines a matrix $n\times n$ Hamiltonians $H_+$ and $H_-$ of Schr\"odinger form according to \gl{splet},
will be called by us {\it complement} for the operator $Q_N^-$ with respect to the Hamiltonian $H_+$. We shall denote the complement for a matrix linear differential intertwining operator $Q_N^-$ as $(Q_N^-)^{\frak C}$, so that the following equality takes place in the conditions of Theorem 5,
\[(Q_N^-)^{\frak C}=Q_{N'}^+.\]

\vskip1pc

{\bf Corollary 9.} If the intertwining operator $Q_N^-$ in the conditions of Theorem 5 is weakly non-minimizable then the intertwining operator $((Q_N^-)^{\frak C})^{\frak C}$ possesses by the following properties:
\renewcommand{\labelenumi}{\rm{(\theenumi)}}
\begin{enumerate}
\item $((Q_N^-)^{\frak C})^{\frak C}=Q_N^-$;
\item any eigenvalue of the matrix $T^+$ is an eigenvalue of the matrix $T^-$ and, conversely, any eigenvalue of the matrix $T^-$ is an eigenvalue of the matrix $T^+$;
\item the highest of the orders of Jordan blocks for the eigenvalue $\lambda_l$ in a normal (Jordan) form of the matrix $T^+$ is equal to the highest of the orders of Jordan blocks for the same eigenvalue in a normal (Jordan) form of the matrix $T^-$, $l=1$, \dots, $L$.
\end{enumerate}

\vskip1pc

{\bf Corollary 10.} Suppose that
\renewcommand{\labelenumi}{\rm{(\theenumi)}}
\begin{enumerate}
\item the operator $Q_N^-$ in the conditions of Theorem 5 is represented as a product of a matrix $n\times n$ linear differential intertwining operators $P_M^-$ and $K_{N-M}^-$ of the orders $M$ and $N-M$ respectively, $0\leqslant M\leqslant N$, so that there is a matrix $n\times n$ intermediate Hamiltonian $H_M$ of Schr\"odinger form and the equalities \gl{interm2.4} take place;
\item $\varkappa_{l1}$ ($\varkappa_{l2}$) is the maximal of orders of Jordan blocks corresponding to the eigenvalue $\lambda_l$ in a normal (Jordan) form of the matrix $T$ of the operator $P_M^-$ ($K_{N-M}^-$) respectively, $l=1$, \dots, $L$ (if $\lambda_l$ does not belong to the spectrum of the matrix $T$ of the operator $P_M^-$ ($K_{N-M}^-$) then  $\varkappa_{l1}=0$ ($\varkappa_{l2}=0$) respectively by definition, $l=1$, \dots, $L$).
\end{enumerate}
Let us note some facts before the main results of this corollary.
\renewcommand{\labelenumi}{\rm{(\theenumi)}}
\begin{enumerate}
\item Any chain from a canonical basis in $\ker Q_N^-$ of $\varkappa_l$ associated vector-functions of $H_+$ for the spectral value $\lambda_l$ can be divided into two parts (see the Sec. 2.3 of \cite{sokolov15}): chain of $\varkappa'_{l1}$ associated vector-functions that belong to $\ker P_M^-$ and set of associated vector-functions which are mapped by $P_M^-$ into a chain of $\varkappa'_{l2}=\varkappa_l-\varkappa'_{l1}$ associated vector-functions of $H_M$ for the same spectral value $\lambda_l$, $l=1$, \dots, $L$.
It is evident that $\varkappa'_{l1}\leqslant\varkappa_{l1}$ and $\varkappa'_{l2}\leqslant\varkappa_{l2}$, $l=1$, \dots, $L$. Hence,
\[\varkappa_{l1}+\varkappa_{l2}\geqslant\varkappa_l,\qquad l=1,\ldots,L.\la{ner10}\]
\item Any eigenvalue of the matrix $T$ of $P_M^-$ belongs to the spectrum of the matrix $T^+$ since any element of a canonical basis in $\ker P_M^-$ can be represented (as an element of $\ker Q_N^-$) in the form of linear combination of elements of a canonical basis in $\ker Q_N^-$ and this linear combination cannot contain elements corresponding to different eigenvalues of $T^+$.
\item Any element of a canonical basis in $\ker K_{N-M}^-$ can be represented in the form $P_M^-\Phi$ with some vector-function $\Phi(x)$. This vector-function as element of $\ker Q_N^-$ can be represented in the form of linear combination of elements of a canonical basis in $\ker Q_N^-$. This linear combination after mapping by $P_M^-$ in view of the Sec. 2.3 of \cite{sokolov15} transforms into linear combination of associated vector-functions of $H_M$ for one of eigenvalues of $T^+$. Hence, any eigenvalue of the matrix $T$ of $K_{N-M}^-$ belongs to the  spectrum of the matrix $T^+$.
\end{enumerate}
Taking these facts and Theorem 5 into account, we obtain that
the following equality is valid,
\[(P_M^-)^{\frak C}(K_{N-M}^-)^{\frak C}=\Big[\prod_{l=1}^L(H_+-\lambda_lI_n)^{\varkappa_{l1}+\varkappa_{l2}-\varkappa_l}\Big](K_{N-M}^-P_M^-)^{\frak C},\]
wherefrom it follows that
\[(K_{N-M}^-P_M^-)^{\frak C}=(P_M^-)^{\frak C}(K_{N-M}^-)^{\frak C}\quad\Leftrightarrow\quad\varkappa_{l1}+\varkappa_{l2}=\varkappa_l,\quad\forall\,  l=1,\ldots,L\]
and that $(K_{N-M}^-P_M^-)^{\frak C}$ is the result of weak minimization of $(P_M^-)^{\frak C}(K_{N-M}^-)^{\frak C}$.\\

{\bf Remark 9.} In the simplest case $n=2$, $N=1$ there are three possibilities: (1) $L=1$, $g_1^-=2$; (2) $L=1$, $g_1^-=1$ and (3) $L=2$. In accordance with Theorem 5 if $L=1$, $g_1^-=2$ then $\varkappa_1=1$ and the order $N'=1$, if  $L=1$, $g_1^-=1$ then $\varkappa_1=2$ and the order $N'=3$ and if $L=2$ then $\varkappa_1=\varkappa_2=1$ and the order $N'=3$ as well. The following example clarifies the construction of the intertwining operator $Q_3^+$ in the subcase $L=2$ and reasons of absence for the intertwining operator $Q_1^-$ in this subcase an operator of the first order that intertwines the same Hamiltonians in the opposite direction and such that its products with $Q_1^-$ are polynomials of the corresponding Hamiltonians with number (not matrix) coefficients. Let us consider two pairs of  scalar Hamiltonians, intertwined by operators of the first order,
\[h_i^\pm=q_{1i}^\pm q_{1i}^\mp+\lambda_i,\qquad q_{1i}^\pm h_i^\mp=h_i^\pm q_{1i}^\pm,\qquad q_{1i}^\pm=\mp\partial+\chi_i(x),\qquad i=1,2,\]
where $\lambda_1$, $\lambda_2$ and $\chi_1(x)$, $\chi_2(x)$ are arbitrary numbers and sufficiently smooth functions respectively. Then the matrix Hamiltonians
\[H_+=\begin{pmatrix}h_1^+&0\\0&h_2^+\end{pmatrix},\qquad H_-=\begin{pmatrix}h_1^-&0\\0&h_2^-\end{pmatrix}\]
are intertwined by the operators
\[Q_1^+=\begin{pmatrix}q_{11}^+&0\\0&q_{12}^+\end{pmatrix},\qquad Q_1^-=\begin{pmatrix}q_{11}^-&0\\0&q_{12}^-\end{pmatrix},\]
so that
\[Q_1^+H_-=H_+Q_1^+,\qquad Q_1^-H_+=H_-Q_1^-,\]
but the products of these operators
\begin{equation}Q_1^+Q_1^-=H_+-\begin{pmatrix}\lambda_1&0\\0&\lambda_2\end{pmatrix},\qquad Q_1^-Q_1^+=H_--\begin{pmatrix}\lambda_1&0\\0&\lambda_2\end{pmatrix}\la{prod21}\end{equation}
for $\lambda_2\ne\lambda_1$ are not polynomials of the corresponding Hamiltonians with number coefficients (the coefficients at $\partial^0$ are not proportional to the identity matrix $I_2$). The intertwining operator $Q_3^+$ takes in this case the form
\[Q_3^+=\begin{pmatrix}(h_1^+-\lambda_2)q_{11}^+&0\\0&(h_2^+-\lambda_1)q_{12}^+\end{pmatrix},\qquad Q_3^+H_-=H_+Q_3^-\]
and the products
\[Q_3^+Q_1^-=(H_+-\lambda_1I_2)(H_+-\lambda_2I_2),\qquad Q_1^-Q_3^+=(H_--\lambda_1I_2)(H_--\lambda_2I_2)\]
are polynomials of the corresponding Hamiltonians with number coefficients. For $\lambda_2=\lambda_1$ the operator $Q_3^+$ is weakly minimizable,
\[Q_3^+=(H_+-\lambda_1I_2)Q_1^+,\qquad \lambda_2=\lambda_1,\]
and the products (\ref{prod21}) are polynomials of the corresponding Hamiltonians with number coefficients.

\section*{Conclusions}

In conclusion we present some questions and problems for future work. 
\renewcommand{\labelenumi}{\rm{(\theenumi)}}
\begin{enumerate}
\item To investigate  the question on (in)dependence of matrix intertwining operators and on basis of such operators (by analogy with \cite{anso03} in scalar case). 
\item To investigate the properties of nontrivial matrix symmetry operator (by analogy with \cite{anso03,anso09} in scalar case).
\item To investigate different partial cases of intertwining relations (for example, with Hermitian or with symmetric intertwined Hamiltonians etc.).
\item  To investigate and classify different types of irreducible matrix differential intertwining operators (by analogy with \cite{acdi95,anca04,samsonov99,anso06,sokolov07,sokolov10,ferneni00,tr89,dun98,khsu99,fermura03,fermiros02,fermiros02',fersahe03,samsonov06} in scalar case).
\item To generalize the achieved results to the case of matrix differential intertwining operator with degenerate eldest coefficient.
\end{enumerate}

\section*{Acknowledgment}

The author is grateful to  A.A. Andrianov for critical reading of the paper and valuable and inspiring discussions and to M.V. Ioffe and S.S. Afonin for interest to the work and for support. This paper was supported by RFBR grant 18-02-00264-a.

\end{document}